\documentstyle[11pt,newpasp,twoside,epsf]{article}
\reversemarginpar

\begin{document}

\title{Radio Observations of Nearby HST BL Lacs}
 \author{Marcello Giroletti \& Gabriele Giovannini}
\affil{Istituto di Radioastronomia, via Gobetti 101, Bologna, Italy}
\affil{Dip. di Astronomia dell'Univ. di Bologna, via
 Ranzani 1, Bologna, Italy}
\author{Greg B. Taylor}
\affil{National Radio Astronomy Observatory, Socorro, NM~87801, USA}
\author{Renato Falomo}
\affil{Oss. Astr. di Padova, vicolo Osservatorio 5, Padova, Italy}

\begin{abstract}
We present a radio/optical study of a sample of BL Lac objects at
arcsecond and milliarcsecond resolution. The sample consists of 30
nearby ($z<0.2$) BL Lacs, mostly X-ray selected.  These nearby objects
are weak in the radio, yet we have successfully observed them with the
VLA, VLBA and EVN, revealing interesting morphologies. Most
importantly, we find evidence of relativistic boosting even in these
low power sources (dominant core, one-sided jets). We present a
radio/optical comparison including HST observations which have
detected compact optical cores in all objects.
\end{abstract}

\section{Introduction}

In spite of much success and major achievements, the unification of FR
I radio galaxies and BL Lac type objects is still an open issue,
requiring further studies. In particular, a study of a sample of
nearby BL Lacs can be informative. In fact, considering the closest
objects allows us to study also the faintest sources, which is
important in many respects.

First, the weak end of the luminosity function has been poorly studied
in the radio at high resolution. The only extensive study on a
complete sample of BL Lacs is based on objects of any redshift,
brighter than 1 Jy at 1.4 GHz (Stickel et al. 1991). This excludes the
weakest objects in the radio, i.e. those objects for which the low
luminosity peak in the SED rises more slowly, peaking at X-ray (High
frequency peaked BL Lacs, HBL) rather then optical/UV (Low frequency
peaked BL Lacs, LBL). This objects are less well known and somehow
more extreme; for example, they are the best candidates for the
presence of TeV emission. Last but not least, in accordance with the
subject of this conference, the milliarcsecond resolution obtainable
with VLBI makes it possible to investigate an extraordinarily fine
linear resolution (1~mas $\sim$ 2~pc at $z = 0.1$,
H$_0=65$\,km\,s$^{-1}$\,Mpc$^{-1}$) providing a wealth of information
at centimeter wavelengths.

For these reasons, we are considering a sample of 30 objects, selected
from a large set of 110 BL Lac objects with the condition of
$z<0.2$. High resolution observations in the optical are presented in
Falomo et al. (2000). The average total radio power at 1.4 GHz, as
derived from the NVSS (Condon et al. 1998), is almost two orders of
magnitude lower than in the 1 Jy sample ($\langle {\rm Log} P_{\rm 1.4
GHz} \rangle = 24.8 \pm 0.6$ and $\langle {\rm Log} P_{\rm 1.4 GHz, 1
Jy} \rangle = 26.8 \pm 0.9$, respectively). Some well known objects,
such as BL Lac itself, Mkn 421, 3C\,371 are common to both samples;
however, about 50\% of the objects in our sample have never been
observed with any VLBI array, or even with the VLA. We present here
results for three objects in the sample (Sect. 2), and discuss in the
light of our new observations some of the properties of the sample as
a whole (Sect. 3).

\section{Observations and Images}

We have obtained observing time with the VLA at 1.4 GHz, in A (10
hrs/19 sources) and C (5 hrs/9 sources) configuration, in order to
complete observations on kiloparsec scales for all the sources. The
observations were completed between February and October 2002.

At pc scale resolution, 15 out of the 30 sources in the sample needed
new or higher fidelity maps. We observed all these sources with the
VLBA at 6 cm on 2002 Feb 17, 18, and 19, for about one hour each. More
observations were obtained with the EVN at 18 cm (12 hrs/6 sources on
2002 Jun 7) and the EVN+MERLIN at 6 cm (12 hrs/2 sources on 2002 May
27).

 \begin{figure}
   \plotone{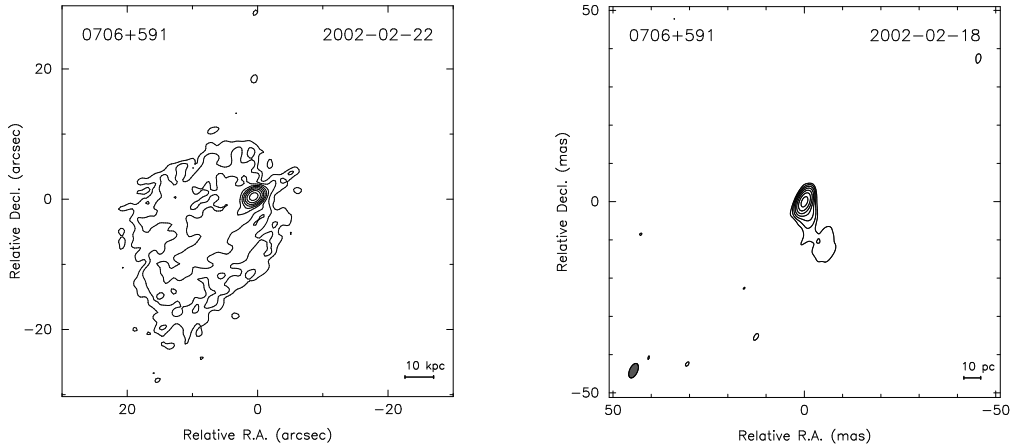}
   \caption{Images of 0706+591. Left: VLA at 1.4 GHz, A configuration
    (HPBW 1.6'' $\times$ 1.2'', PA $-57^\circ$); right: VLBA at 5
    GHz. (HPBW 3.6 $\times$ 1.4 mas, PA $-23^\circ$)}
 \end{figure}

{\bf 0706+591} -- This object is located at $z=0.125$ and is
 classified as HBL. The VLA image in A configuration (left panel in
 Fig. 1), reveals a dominant core and an extended structure oriented
 at PA $\sim 130^\circ$ (measured north to east). The total flux
 density is 161 mJy. The VLBA image at 5 GHz (right panel) has a peak
 of 33 mJy/beam and presents a jet in a direction perpendicular to the
 kpc scale. This is quite common behaviour for BL Lac objects in our
 sample.

 \begin{figure}
   \plotone{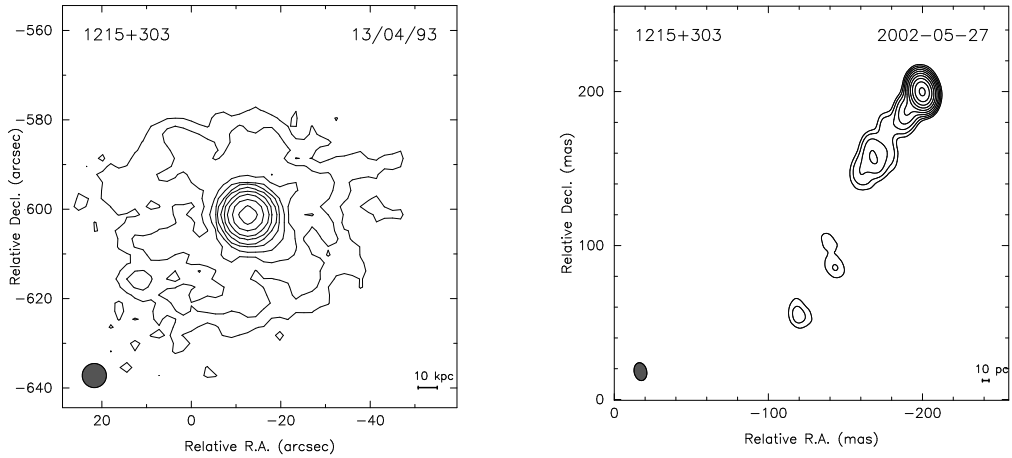} 
   \caption{Images of 1215+303. Left: image from the FIRST survey
    (Becker et al. 1995); right: EVN+MERLIN at 5 GHz.}
 \end{figure}

{\bf 1215+303} -- The HBL 1215+303 is located at $z=0.130$. For this
 object, a good map is available in the FIRST survey (Becker et
 al. 1995), as shown in Fig. 2 (left). A 377 mJy core dominates a
 symmetric halo structure of about 50'' diameter. No preferred
 direction is visible on this scale. For this reason, we performed
 EVN+MERLIN observations at 5 GHz, in order to image the inner
 structure out to the largest possible distance from the inner
 core. The right panel in Fig. 2 shows a one-sided, long, straight
 jet. The large jet/counter-jet ratio ($R\ga 150$) indicates that the
 effects of the beaming are important in this source; indeed, one
 could expect a small angle to the line of sight, if the kpc halo
 structure were interpreted as a lobe seen end-on.

\begin{figure}[b]
   \plotone{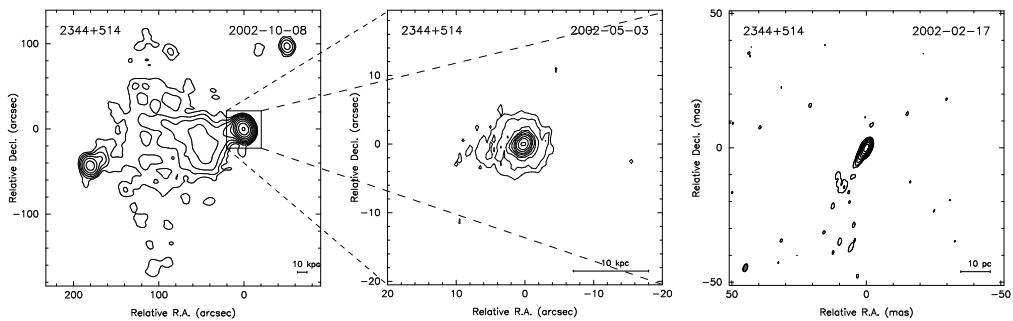}
   \caption{Images of 2344+514. From left to right: VLA at 1.4 GHz, C
    configuration; VLA at 1.4 GHz, A configuration; VLBA at 5 GHz.}
 \end{figure}
 
{\bf 2344+514} -- This is one of the closest objects in the sample
 ($z=0.044$) and presents a puzzling morphology (Fig. 3). It looks
 like a double when observed in C configuration, with faint, extended
 emission connecting the two compact components. A third, misaligned,
 less bright component could be a background object (even if it seems
 aligned with a small extension in the faintest component of the
 double). However, imaging of the strongest component in A
 configuration reveals a core-halo morphology. Finally, the high
 resolution VLBA image shows a $\sim 20$ mas jet almost orthogonal to
 the large scale structure axis. Note that {\it Chandra} observations
 have revealed diffuse X-ray emission in the environment of this
 source (Donato et al. 2003).

\section{Discussion}

\begin{figure}
  \plotone{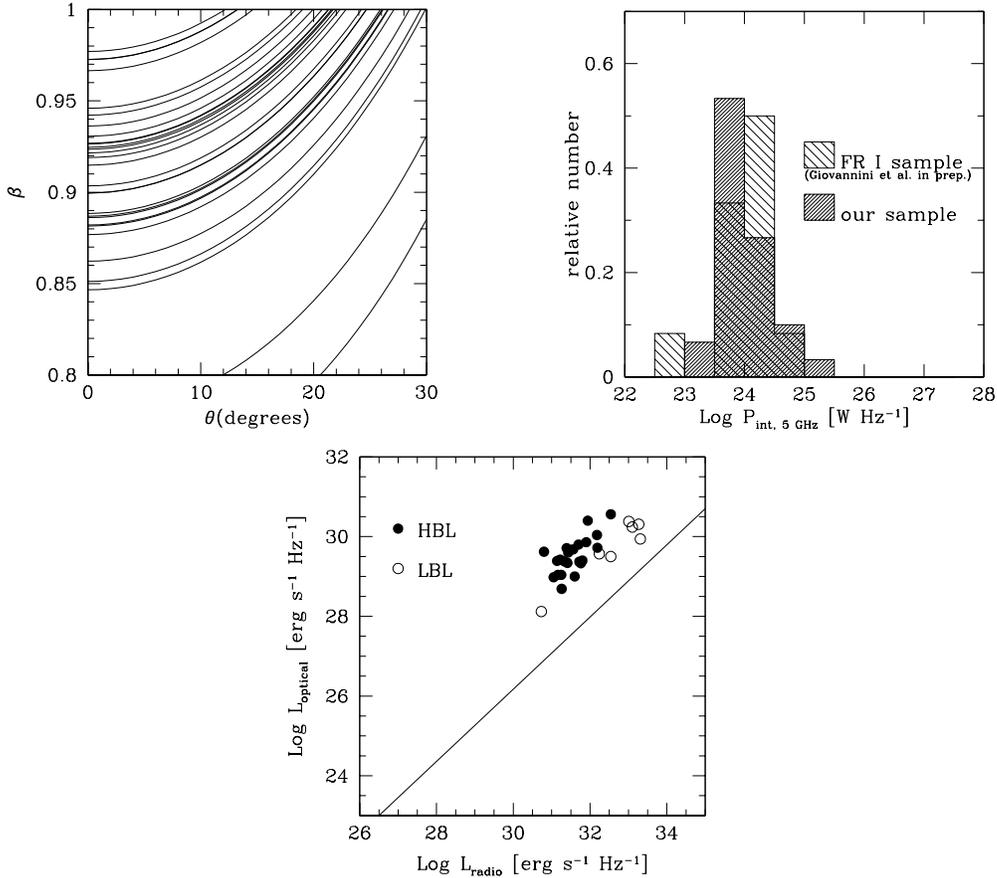}
  \caption{Top left panel: the $(\theta, \beta)$ plane; each curve
   corresponds to one object. Top right: histogram of de-boosted core
   powers for the present sample and FR Is of similar $z$. Bottom:
   Optical vs. radio core luminosities; the straight line represents
   the linear correlation found by Chiaberge et al. (2000) for radio
   galaxies.}
\end{figure}

Our radio data show a variety of behaviours: core-jet, core-halo,
double compacts and even a few head-tails, clearly related to the
membership to galaxy clusters. However, all the sources are
characterized by a compact, bright, dominant radio core in the BL Lac
position. Thanks to the well-known correlation between the core and
total radio power (Giovannini et al. 1988; 2001), we are therefore
able to derive the allowed range of $\beta$ and $\theta$ for all the
sources in the sample (see Fig. 4, left panel). We obtain values in
good agreement with the expectations of the unified scheme; in
particular $\beta \ga 0.9$ (i.e. relativistic velocity) and $\theta
\la 25^\circ$ (small angle of view).

It is not possible to decide which combination of these parameters is
the most realistic and we do not even know if all the objects have the
same intrinsic properties. However, following Giovannini et al. 2001,
we assume a Lorentz factor $\Gamma = 5$ ($\beta=0.98$) and derive for
every single source the corresponding angle $\theta_5$ and Doppler
factor $\delta_5$. Then, we use these results to de-boost the core
power at 5 GHz and obtain intrinsic values. The comparison with a
sample of FR I radio galaxies of similar redshift (Giovannini et al.,
in prep.), shows that the two population overlay well (Fig. 4, right
panel).

Finally, we compare the luminosities of the BL Lac cores in the radio
and in the optical. Figure 4 (at bottom) shows that the luminosities
correlate with a linear law, just as the 3C and B2 radio galaxies do
(straight line). This indicates a common origin for both radio and
optical emission. There is however a considerable offset ($\sim 2$
orders of magnitude) between BL Lacs and radio galaxies, suggestive of
a larger beaming factor for the region emitting at optical frequency
(Chiaberge et al. 2000). In turn, this can be explained by the
presence of a two-velocity regime (fast inner spine and slow external
shear) in pc scale jets and/or by a spine velocity decrease from the
optical to the radio emitting region.

\section{Summary}

We have completed parsec and kiloparsec scale radio observations for
all sources in our sample. All objects show a dominant, compact core
surrounded by different features: jets, halos, other compact
components; bending in jets from pc to kpc scale is often present. We
derive luminosities consistent with a parent population of FR I type
and confirm the existence of a linear correlation between optical and
radio core luminosity.

\end{document}